# Possible coexistence of double-Q magnetic order and chequerboard charge order in the re-entrant tetragonal phase of $Ba_{0.76}K_{0.24}Fe_2As_2$

Jianqiang Hou [a], Chang-woo Cho [a], Junying Shen [a], Pok Man Tam [a], I-Hsuan Kao [a], Mang Hei Gordon Lee [a], Peter Adelmann [b], Thomas Wolf [b] and Rolf Lortz [a][*]

[a] Department of Physics, The Hong Kong University of Science and Technology, Clear Water Bay, Kowloon, Hong Kong, China

[b] Institute for Solid State Physics, Karlsruhe Institute of Technology, PO Box 3640, 76021 Karlsruhe, Germany

**ABSTRACT**

We investigate the re-entrant tetragonal phase in the iron-based superconductor $Ba_{0.76}K_{0.24}Fe_2As_2$ by DC magnetization and thermoelectrical measurements. The reversible magnetization confirms by a thermodynamic method that the spin alignment in the re-entrant $C_4$ phase is out-of-plane, in agreement with an itinerant double-**Q** magnetic order [Allred et al., Nat. Phys. **12**, 493 (2016)]. The Nernst coefficient shows the typical unusually large negative value in the stripe-type spin density wave (SDW) state owing to the Fermi surface reconstruction associated with SDW and nematic order. At the transition into the re-entrant $C_4$ tetragonal phase it hardly changes, which could indicate that instead of a complete vanishing of the associated charge order, the spin reorientation could trigger a redistribution of the charges to form a secondary charge order, e.g. in form of a chequerboard-like pattern that no longer breaks the rotational $C_4$ symmetry.

## 1. Introduction

The superconducting state is characterized by a two-component order parameter as a consequence of the spontaneous $U(1)$ gauge symmetry breaking below its transition temperature $T_c$ [1,2]. The order parameter, which is proportional to the superconducting gap function and the amplitude of the Cooper pair wave-function, determines many physical properties [1-3]. Its symmetry is directly related to the superconducting pairing mechanism. In unconventional superconductors, additional symmetries are broken in addition to $U(1)$, and the superconducting phase usually appears together with other forms of orders, such as magnetism, charge or orbital order [4-6]. A detailed understanding of how these orders are related to superconductivity (e.g. competing or coexisting) is of primary importance to decipher the superconducting mechanism.

[*] Corresponding author. Email: lortz@ust.hk

In iron-based superconductors, there are at least two additional forms of order [4,7-11]. For instance, in the $BaFe_2As_2$ (Ba122) family, the parent compound features an antiferromagnetic stripe-type spin density wave (SDW) order in which chains of parallel spins are adjacent to chains with opposite spin direction [5]. For hole-doped 122 materials, the SDW transition more or less coincides with a spontaneous breaking of rotational symmetry (nematic order) from a tetragonal ($C_4$) symmetry to an orthorhombic structure with $C_2$ symmetry in the low temperature regime, while these two orders set in at slightly different temperatures for electron-doped 122 materials [5]. Both forms of order are gradually suppressed by increasing the dopant concentration until superconductivity emerges [5]. The nematic order is associated with an electronic instability [12-18] that causes pronounced in-plane anisotropy of the Fe-As layers. In addition to a simple structural distortion, candidates are charge or orbital order in the form of a spontaneously occurring different occupancy of the $d_{xz}$ and $d_{yz}$ Fe orbitals, or an itinerant magnetic scenario where a spin order with different static spin susceptibilities develops along the $q_x$ and $q_y$ directions of the Brillouin zone as precursors of the SDW order [18].

The phase diagram of hole doped Ba122 compounds, such as $Ba_{1-x}K_xFe_2As_2$ [19] $Sr_{1-x}Na_xFe_2As_2$ [20] and $Ba_{1-x}Na_xFe_2As_2$ [21] is particularly rich, with at least one additional phase occurring in addition to the orthorhombic stripe-type SDW phase above $T_c$. It has been identified by a distinct magnetic order in which the $C_4$ rotational symmetry is restored [21], and the bulk $T_c$ is greatly reduced when this phase is present [19]. It is unclear whether this transition is driven by magnetic interactions, or rather by a reconstruction of the Fe 3d orbitals [18]. The transition between the orthorhombic stripe-type SDW order and this re-entrant tetragonal phase has been associated with a spin re-orientation to an itinerant double-**Q** spin density wave state [20,22]. In this phase half of the iron sites were reported to be non-magnetic, whereas the other half show a doubling of the magnetic moment. In addition, it was predicted that the double-**Q** magnetic order may be accompanied by a secondary chequerboard charge order in which the non-magnetic sites have a distinct charge [20,23].

In this paper, we investigate a $Ba_{0.76}K_{0.24}Fe_2As_2$ single crystal with various experimental probes including DC magnetization, thermal expansion and thermoelectric transport in the vicinity of the re-entrant tetragonal phase. The thermal expansion data confirms the presence of the re-entrant tetragonal phase in form of the orthorhombic distortion within the $ab$ plane. Detailed DC magnetization data confirm in the reversible regime by a bulk thermodynamic method that the spin alignment is out-of-plane in the re-entrant $C_4$ phase [18,20,22], while the Nernst effect shows a large negative signal as a signature of the Fermi surface reconstruction in the stripe-type SDW phase, which is unchanged when the temperature is lowered across the transition into the re-entrant $C_4$ phase. The Nernst coefficient has been shown to be particularly sensitive to nematic order [24]. This provides evidence that the double-**Q** magnetic order is indeed accompanied by a distinct charge or orbital order [20], which no longer breaks the $C_4$ rotational symmetry of the material.

## 2. Experimental details

The $Ba_{0.76}K_{0.24}Fe_2As_2$ single crystal used in this study was grown from self-flux in an $Al_2O_3$ crucible. Ba and K were mixed with pre-reacted FeAs in the desired ratio and placed into the crucible. The crucible was sealed in a steel container and heated to 1151 ºC. The crucible was then cooled down very slowly to 1051 ºC at 0.2-0.5 ºC/h. In order to decant the remaining flux, the crucible was tilted at the end of the growth process and slowly pulled out of the furnace. The slow cooling rate allowed us to grow large single crystals of high homogeneity with 4-8 mm length in the *ab* plane, which is crucial for high-resolution thermal expansion measurement. The exact K content x of the samples typically differs from the starting composition, and was precisely determined by energy-dispersive x-ray analysis (EDX) and by four-circle x-ray diffraction (XRD). All experiments were performed on the same single crystal, which is from the batch of samples used in Ref. 19, in which further detailed characterization can be found.

The DC magnetization was measured with a Quantum Design Vibrating Sample SQUID magnetometer under both field cooled and zero field cooled conditions, which provides an absolute resolution exceeding $10^{-8}$ emu. The thermal expansion was measured in a homemade capacitance dilatometer in which a sample is pushed against a cantilever by a fine screw mechanism. The cantilever forms one of the two plates of a capacitor. The legs of the cantilever are separated by a thin sapphire disk. With this design, we can measure the thermal expansion in a very compact set-up in a highly reproducible manner. The capacitance of the dilatometer was measured with a General Radio capacitance bridge in combination with a digital Stanford Research SR830 lock-in amplifier, which provides an absolute resolution for length changes $\Delta L$ of the sample exceeding $10^{-3}$ ångstrom. Data was taken during temperature sweeps with constant sweep rate of 1 K/min, and used to confirm the existence of transition into the reentrant tetragonal phase, since the linear thermal expansion measured in plane directly monitors the vanishing of the orthorhombic distortion. The thermal conductivity and the thermoelectric Nernst and Seebeck coefficients were measured by attaching one end of the sample in form of a ~1 cm long thin slab to a copper heat sink, while a small ~100 Ohm resistor was attached to the other end. The thermal gradient was determined with two Au99.93/Fe0.07 - Chromel thermocouples, which was carefully calibrated in magnetic fields in a preceding separate experiment. The Nernst and Seebeck voltages were measured with help of four electrical contacts that allowed us to probe the longitudinal and transverse thermoelectrical voltages. The Nernst effect was determined in a magnetic field of 6 T applied perpendicular to the FeAs layers. In order to eliminate spurious longitudinal components from a non-ideal electrode arrangement in the Nernst measurements, we reversed the field direction and repeated the experiment under otherwise identical conditions. This allowed us to separate the longitudinal and transverse voltage components due to their different behavior with respect to a field reversal.

## 3. Results and Discussion

Fig. 1 shows the thermoelectric Seebeck coefficient in a magnetic field of 0 and 6 T applied perpendicular to the *ab* plane (Fig. 1a), the thermal conductivity (Fig. 1b), and the linear thermal expansion $\Delta L_{ab}/L_{ab}$, all measured within the partially twinned *ab* plane of $Ba_{0.76}K_{0.24}Fe_2As_2$ (Fig. 1c). All data is plotted over the same temperature range for comparison.

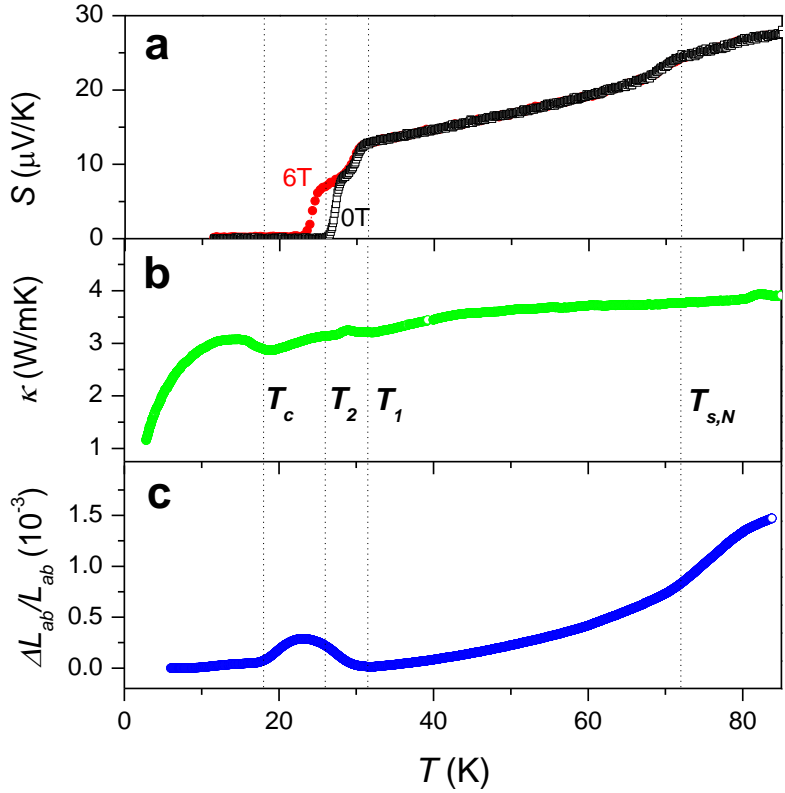

**FIG. 1.** (a) Thermoelectric Seebeck coefficient in 0 (open squares) and 6 T (closed circles) of $Ba_{0.76}K_{0.24}Fe_2As_2$. (b) Zero field thermal conductivity. (c) Linear relative thermal expansion $\Delta L_{ab}/L_{ab}$ measured along one crystalline direction within the *ab* plane. Characteristic temperatures $T_{s,N}$, $T_1$, $T_2$ and $T_c$ are marked by the vertical dotted lines (see text for details).

Four characteristic temperatures can be identified from the data: At $T_{s,N} \sim 72$ K the thermal expansion and the Seebeck coefficient display anomalies that correspond to the onset of the stripe-type SDW and nematic order [19]. The anomaly in the thermal expansion is a measure of the orthorhombic distortion that develops as a consequence of the nematic order, and visible here because the sample is pushed with a spring mechanism along one crystalline in-plane direction against a movable capacitor plate. The associated force partially prevents a twinning of the sample, which otherwise is known to occur in this phase [19]. At $T_1 \sim 32$ K, the Seebeck coefficient, which is closely related to the electrical resistivity, shows a first sharp drop, the thermal conductivity displays a step like anomaly, and the orthorhombic distortion in the thermal expansion is gradually removed. Below $T_2 \sim 27$ K, a second sharp drop to zero value occurs in the Seebeck coefficient. No sharp anomaly is visible in any of the other quantities at this temperature. Below $T_c \sim 18$ K, finally the orthorhombic distortion in $\Delta L_{ab}/L_{ab}$ appears to become almost fully restored. The latter has been identified as the true bulk $T_c$ as evidenced from previous thermal expansion and specific heat measurements [19]. This is further confirmed by the characteristic superconducting anomaly at $T_c$ in the thermal conductivity, which can be considered as a bulk probe closely related to the specific heat. The drop in the Seebeck coefficient below $T_2$ thus is caused by filamentary superconductivity, which likely is associated with domain walls between tetragonal and orthorhombic regions which may coexist in the temperature range between $T_c$ and $T_2$ [22].

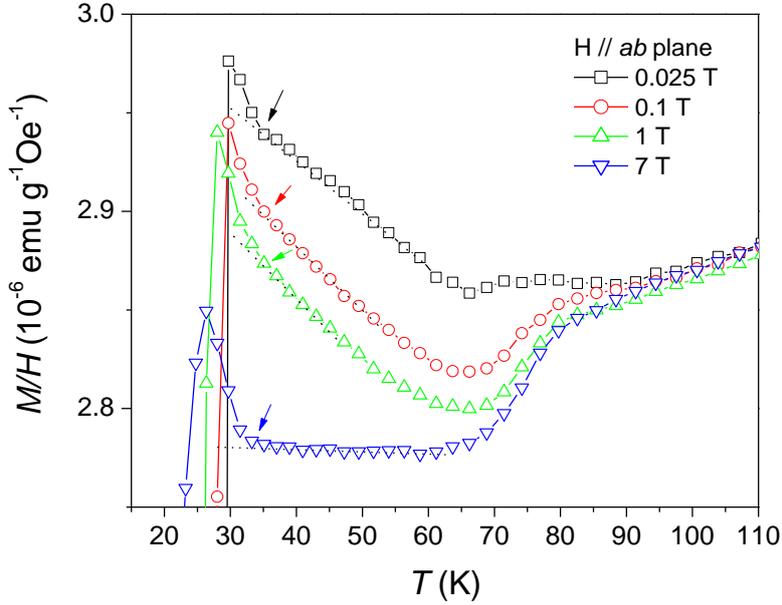

**FIG. 2.** DC Magnetization normalized by the applied magnetic field of $Ba_{0.76}K_{0.24}Fe_2As_2$ in zero field cooled condition with different applied magnetic field parallel to *ab* plane. Arrows marked the transition anomalies at $T_2$.

In Fig. 2, the magnetization of $Ba_{0.76}K_{0.24}Fe_2As_2$ in different magnetic fields applied parallel to the *ab* plane up to 7 T is shown. Note that our interest here is on the normal state above the onset of superconductivity, and we focus on a rather high magnetic field range in which the Meissner signal is weak. With our high resolution magnetization data we can identify the two transition anomalies at $T_{s,N}$ and $T_1$ above the strong drop associated with the onset of filamentary superconductivity at $T_2$. With applied field parallel to *ab* plane, the magnetization data in different fields all merge together above ~80 K and then start to deviate below this temperature. At $T_1$, a kink is visible in all date followed by a distinct upturn of the magnetic moment at lower temperatures, as marked by the arrows.

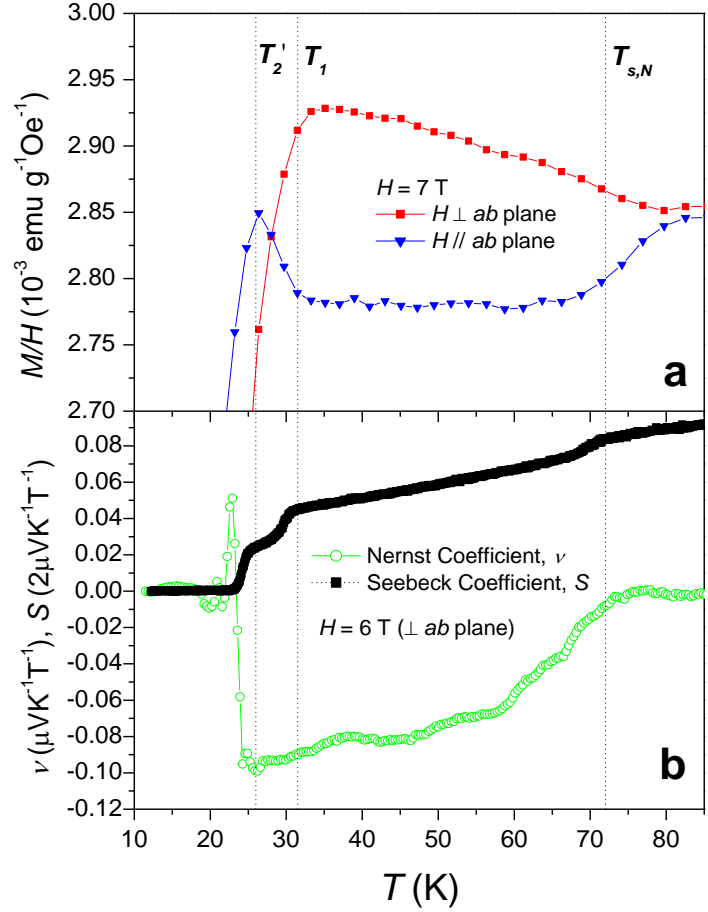

**FIG. 3.** (a) DC Magnetization normalized by the applied magnetic field of $Ba_{0.76}K_{0.24}Fe_2As_2$ in a 7 T magnetic field applied parallel and perpendicular to the *ab* plane. (b) Temperature dependence of the Nernst coefficient $v = N/B$ in comparison to the Seebeck coefficient in a 6 T magnetic field applied perpendicular to the *ab* plane. $T_2'$ corresponds to the characteristic temperature $T_2$, but in an applied magnetic field of 6 T.

Fig. 3 shows the magnetization of $Ba_{0.76}K_{0.24}Fe_2As_2$ in an applied magnetic field of 7 T parallel and perpendicular to *ab* plane (a), together with the thermoelectric Nernst and Seebeck coefficient (b) on the same temperature scale for comparison. The magnetization data measured under these two conditions shows a pronounced anisotropy below $T_{s,N}$ due to the onset of the stripe-type antiferromagnetic order. While the data with field applied perpendicular to the *ab* plane increase below $T_{s,N}$ first and then drop abruptly below $T_1$, the data for field applied parallel to the *ab* plane show the opposite behavior: the magnetization drops below $T_{s,N}$ and then abruptly increase below $T_1$ before both data show the drop due to the onset of filamentary superconductivity below $T_2'$. The latter characteristic temperature corresponds to the phase transition at $T_2$ which is somewhat lowered here by the strong applied magnetic field. All data shown here was taken under ZFC condition. No difference between ZFC and FC conditions was found above $T_2'$.

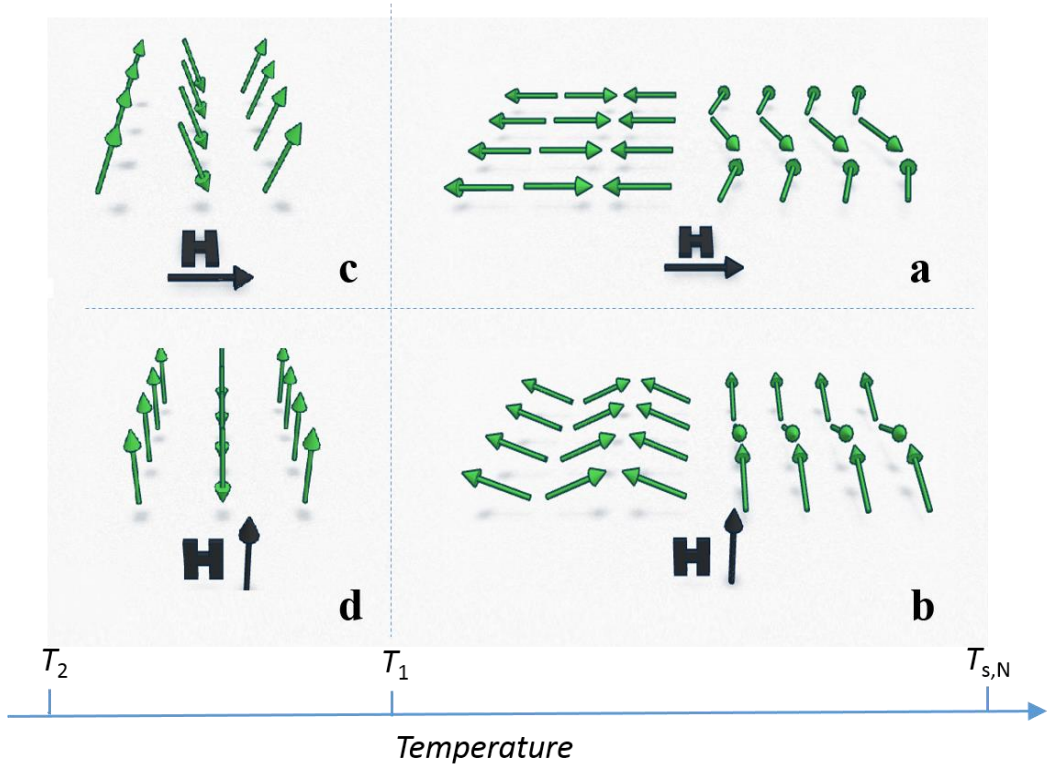

**FIG. 4.** Effect of a magnetic field applied parallel (a, c) and perpendicular (b, d) to the *ab* plane of $Ba_{1-x}K_xFe_2As_2$ in the stripe-type SDW phase with in-plane antiferromagnetic order (a, b), and in the re-entrant tetragonal order (c, d) with an assumed out-of-plane spin arrangement.

The high magnetic field value is here chosen as a tool to probe how easily the spins can be tilted along the field direction. Below $T_{s,N}$ the spins align in form of the stripe-type anti-ferromagnetic order within the *ab* plane. Due to the twinning, there will be domains in which the spins point perpendicular to the field, and in which they point along the field direction. While the spins in the twin domains perpendicular to the field may be slightly tilted by the strong field into the field direction, the spins in the domains with parallel alignment will be uninfluenced by the field, since a much higher field would be required to flip them all parallel into a field-polarized state. This is illustrated in Fig. 4a and explains the lower magnetization value for the in-plane field orientation in the stripe-type SDW state below $T_{s,N}$. For the field direction perpendicular to the *ab* plane, the field is always applied perpendicular to the spins (Fig. 4b), and the larger magnetization value here indicates that the field can rather easily tilts the spins out of plane.

Below $T_1$ the sample then enters the re-entrant tetragonal phase and the magnetization strongly increases for a 7 T magnetic field applied in-plane, while it decreases for the out-of-plane direction. This behavior is hard to explain within any scenario in which the spins are aligned in-plane, but can be well understood if the spins are oriented out of plane instead, as illustrated in Fig. 4c and d. If the field is applied parallel to the *ab* plane, it will be perpendicular to the spins orientated out-of-plane, and can rather easily tilt the spins along the field direction, thus resulting in an increase of the magnetization value. For the out-of-plane, the field would need to flip all the antiparallel spins in order to achieve a sizable change in the magnetization value by entering the field polarized state, which would require a much higher magnetic field. Therefore, a decrease of the magnetization thus confirms that the spin alignment in the re-entrant tetragonal phase is indeed an antiferromagnetic out-of-plane order, which perfectly agrees with the suggested double-**Q** magnetic order [20,23] .

The thermoelectrical Nernst effect has proved to be a powerful tool for understanding the phase diagram of superconductors [24-29]. While the Nernst coefficient is usually very small for ordinary metals in their normal state, a large positive contribution from vortex motion results in type-II superconductors in addition to the regular normal state contribution, and has been used to monitor the presence of superconducting fluctuations above $T_c$ [28]. In addition, am unusually large Nernst coefficient is often observed in presence of nematic order arising from electronic correlations that spontaneously break the rotational symmetry, as observed e.g. in cuprates [25] and iron based superconductors [24,26,27,29].

Fig. 3b shows the thermoelectric Nernst in comparison with the Seebeck coefficient in a 6 T field. The Nernst coefficient is almost zero in the normal non-magnetic state above $T_{s,N}$, but develops a large negative value below. The most interesting part is at $T_1$, where the first drop occurs in the Seebeck coefficient upon entering the re-entrant tetragonal phase. The negative Nernst coefficient value generally associated with the nematic and SDW order [24,26,27,29] remains unchanged within the experimental resolution, until it abruptly crosses zero and forms a positive peak below $T_2'$. The latter originates presumably from a positive vortex motion contribution [28] in small superconducting domains, before it goes back to zero in the vicinity of the bulk $T_c$ value, below which the vortex contribution vanishes due to the onset of strong flux pinning in the bulk superconducting state.

In an ordinary one-band metal the Nernst coefficient is typically very small as a result of the `Sondheimer cancellation' [30],

$$\nu = \left( \frac{\alpha_{xy}}{\sigma} - S \tan \theta \right) \frac{1}{B}$$

where $\alpha$ is the Peltier coefficient tensor, $\sigma$ is the longitudinal electrical conductivity, $S$ is the Seebeck coefficient and $\theta$ is the Hall angle. In absence of strong electronic correlations the two terms usually cancel resulting in a vanishing of the Nernst coefficient $\nu$. A large Nernst coefficient may thus likely indicate the presence of strongly correlated electronic states, such as superconductivity, nematic order or SDW states [24-29]. For example in LaFeAsO, the violation of the Sondheimer cancellation in the nematic state has been explained by a Fermi surface reconstruction with spontaneous breaking of rotational symmetry associated with the nematic order [24,26].

Our observation that the negative value of the Nernst coefficient remains unchanged at $T_1$ suggests that the electronic orders undergo little change at the phase transition between the stripe-type SDW order and the re-entrant tetragonal phase. This is in contradiction with our previous experiments in which we induced the re-entrant tetragonal phase in a more underdoped sample by applying pressure [27]. In the pressure experiment, we observed a very similar negative Nernst coefficient in the nematic stripe-type SDW phase below $T_{s,N}$, but at $T_1$ the Nernst coefficient abruptly changed sign and reached a small positive value in the re-entrant phase. This was a clear sign that not only the spins re-orient themselves, but also the nematic order disappears, which naturally explains the re-entrance of tetragonal order. In the present ambient pressure experiment, the nematic order causing the orthorhombic distortion, must obviously also disappear at $T_1$ to take into account the re-entrant tetragonal crystal symmetry. However, if the underlying charge order would completely vanish at the temperature at which the stripe-type SDW order transforms into the double-$\mathbf{Q}$ SDW order with out-of-plane spin arrangement [20], a significant change of the anomalous Nernst contribution would be expected to occur in a similar manner as in our previous pressure experiment. The fact that no change occurs may thus indicate that the charge order associated with the nematic order does not vanish at $T_1$, but the charges rather rearrange in a way to please the new magnetic double-$\mathbf{Q}$ order. This only agrees with a tetragonal structure if the charges reorient themselves among the $d_{xz}$ and $d_{yz}$ Fe orbitals to form a new type of charge order with modulation along both in-plane directions. This could be likely realized in form of the chequerboard orbital order that has been predicted in Ref. 20. The comparison of the ambient pressure data in this work with our previously reported pressure data [27] thus indicates that the re-entrant tetragonal phases are slightly different at ambient pressure and under pressure. The magnetic double-$\mathbf{Q}$ order may be established in both cases, but while the re-entrant tetragonal phase under pressure appears to be free of charge or orbital order, such order is likely present in a form that agrees with the tetragonal symmetry of the re-entrant phase.

## 4. Conclusions

In summary, we investigated the re-entrant tetragonal phase in the iron-based superconductor $Ba_{0.76}K_{0.24}Fe_2As_2$ with various experimental probes, including DC magnetization, thermal expansion, thermal conductivity and thermoelectrical measurements. The sample shows the transition into the stripe-type SDW state with nematic order at $T_{s,N}$ ~72 K, as evidenced by the thermal expansion data, and the transition into the re-entrant tetragonal phase at $T_1 = 32$ K. Below $T_2 = 27$ K, traces of filamentary superconductivity, presumably in the vicinity of grain boundaries to orthorhombic minority domains, cause a dramatic drop in the Seebeck coefficient, well above the bulk $T_c = 18$ K as evidenced from the thermal conductivity and thermal expansion. The two structural and magnetic transitions at $T_{s,N}$ and $T_1$ are also visible in the magnetization, and by comparing 7 T high field data for field orientation parallel and perpendicular to the FeAs layers, we are able to confirm that the spins reorient themselves from an in-plane stripe-type antiferromagnetic arrangement in the high temperature SDW phase to an out-of-plane arrangement within the re-entrant tetragonal phase, which has been proposed to exhibit a double-**Q** itinerant magnetic order [20]. The unnoticeable change in the Nernst coefficient may indicate that the charge order, which is associated with electronic nematic order in the high temperature stripe-type SDW state, does not vanish at this transition, but undergoes a transformation from nematic order to a chequerboard–type charge order, which causes the orthorhombic distortion to vanish.

## Acknowledgments

We thank U. Lampe and G. Suen for technical support. This paper was supported by grants from the Research Grants Council of the Hong Kong Special Administrative Region, China (SBI15SC10).